\documentclass[aps,prx,reprint]{revtex4-2}

\usepackage{epsfig,graphics}
\usepackage{amsfonts,amssymb,amsmath} 
\usepackage{amsthm}         
\usepackage{color}
\usepackage{makecell}
\usepackage{tabularx}

\theoremstyle{plain}
\newtheorem{theorem}{Theorem}

\begin{document}
\title{Orbital Angular Momentum Experimental Bound on the Maximum Predictive Power of Physical Theories in Multi-Dimensional Systems}

\author{Jianqi Sheng$^1$}

\author{Dongkai Zhang$^{1, 2}$}
\email[]{zhangdk@hqu.edu.cn}

\author{Lixiang Chen$^1$}
\email[]{chenlx@xmu.edu.cn}

\affiliation{$^1$Department of Physics, Xiamen University, Xiamen 361005, China}
\affiliation{$^2$College of Information Science and Engineering, Fujian Provincial Key Laboratory of Light Propagation and Transformation, Huaqiao University, Xiamen 361021, China}

\begin{abstract}
The completeness of quantum mechanics in predictive power is a central question in its foundational study. While most investigations focus on two-dimensional systems, high-dimensional systems are more general and widely applicable. 
Building on the non-extensibility theorem by Colbeck and Renner [Phys. Rev. Lett. 101, 050403 (2008)], which established that no higher theory can enhance the predictive power of quantum mechanics for two-dimensional systems, we extend this result to arbitrarily dimensional systems. 
We connect maximum potential predictive power achievable by any alternative theory to experimentally observable correlations, and establish optimal experimental bounds across varying dimensions by exploiting two-photon orbital angular momentum entangled states with entanglement concentration. 
These bounds falsify a broader class of alternative theories, including Bell's and Leggett's models, and those that remain theoretically ambiguous or experimentally unverified. Our findings not only deepen the foundational understanding of quantum mechanics but also hold significant potential for high-dimensional quantum cryptography. 
\end{abstract}

\maketitle
\section{INTRODUCTION}

The fundamental discrepancy between quantum theory and classical theory lies in their predictive frameworks, with quantum theory being intrinsically probabilistic and classical theory strictly deterministic. This divergence raises a critical question regarding the optimality of quantum theory in predicting measurement outcomes, and whether potential extensions could enhance its predictive accuracy. 
This question has sparked a long-standing debate, tracing back to the seminal paper by Einstein, Podolsky, and Rosen \cite{EPR}, which suggested that the nonlocal correlations observed in quantum mechanics could imply the theory's incompleteness. Their hypothesis introduced the concept of "hidden variables" as a potential explanation for these correlations. 
Several years later, Bell showed that quantum theory's predictions are incompatible with any theory based on the principles of locality and realism \cite{bell1964einstein}. He demonstrated that certain correlations between entangled particles cannot be explained by local hidden variable theories, a conclusion that has been consistently supported by increasingly precise experiments \cite{giustina2015significant,shalm2015strong}. 
The debate over hidden variable models was reignited by Leggett's work \cite{leggett}, which inspired a new wave of experimental investigations of nonlocal realism \cite{groblacher2007experimental,paterek2007experimental,branciard2007experimental,branciard2008testing}. In Leggett's model, the hidden variables are assumed to allow partially nonlocal components, which he demonstrated to be incompatible with quantum theory. 
These arguments aimed to challenge the optimality of quantum theory in predicting measurement outcomes by proposing alternative theories based on hidden parameters that could provide superior predictive power.

Colbeck and Renner extended the discussion of hidden variable theories more broadly from an information-theoretic perspective. 
They concluded that, under the assumption of freely chosen measurement settings, no physical theory can surpass the predictive power of quantum mechanics \cite{Colbeck,colbeck2011no,colbeck2015completeness}. This result can be viewed as a non-extensibility theorem, a generalization of Bell's theorem. 
By utilizing biphoton polarization entanglements, Stuart and coworkers provided solid experimental evidence supporting the non-extendibility theorem \cite{stuart2012experimental}. 
Based on their experimental data, a nontrivial bound was established on the potential increase in predictive power that any alternative theory could achieve. This bound can serve to falsify a class of hidden variable models, including Bell's and Leggett's models \cite{leggett}. 
Note that these results are limited to two-dimensional cases, and the application to higher-dimensional systems remains both theoretically and experimentally unexplored. 

In this work, we extend the theoretical framework of Colbeck and Renner \cite{Colbeck} to arbitrary dimensions and determine optimal experimental bounds across varying dimensions by exploiting two-photon orbital angular momentum (OAM) entanglement. 
The parper is organized as follows. 
First, we connect maximum potential predictive power achievable by any alternative theory in arbitrary dimensions to experimentally observable correlations. 
Next, we experimentally test the correlations by utilizing two-photon orbital angular momentum entangled states with entanglement concentration. 
Finally, we present experimental data across various dimensional systems, establishing bounds on the potential increase in predictive power that any alternative theory could achieve. 
In this way, we establish a crucial connection between experimentally measurable quantities and the maximum predictive power achievable by any alternative theory, providing bounds based on realistic experiments. Notably, these bounds enable the falsification of a broader class of physical theories that provide significantly higher predictive accuracy than quantum theory, including those that remain theoretically ambiguous or experimentally unverified. 

\section{THE NON-EXTENDIBILITY THEOREM FOR ARBITRARY DIMENSIONS}

Before presenting the experiment, we introduce the theoretical framework of Colbeck and Renner's theorem and extend it to arbitrary dimensional systems. 
The scenario under consideration involves a bipartite setup where a source emits two particles directed to two spatially separated detectors controlled by two parties, Alice and Bob. Each party randomly selects a measurement from \(N\) possible choices, denoted by \(A\) and \(B\), where \(A, B\in \{1, \ldots, N\}\). The detectors then produce \(d\)-dimensional outcomes \(X\) and \(Y\), respectively, where \(X,Y\in \{0, \ldots, d-1\}\). 
Now consider potential alternatives to quantum theory that may provide a different description of the measurement process, where the additional information introduced is accessed by choosing an observable \(C\) and obtaining an outcome \(Z\). 
This can be modeled by introducing an additional third party, giving rise to a joint distribution \(P_{XY Z \mid A B C}\) \cite{colbeck2011no}. 
Under the assumption that the measurements can be chosen freely, the joint distribution \(P_{XY Z \mid A B C}\) must satisfies the following non-signaling conditions 
\begin{align}
P_{X Y \mid A B C} & =P_{X Y \mid A B}, \label{XY}\\
P_{X Z \mid A B C} & =P_{X Z \mid A C} ,\label{XZ}\\
P_{Y Z \mid A B C} & =P_{Y Z \mid B C}.\label{YZ}
\end{align}
The proof can be found in \cite{colbeck2011no}. These conditions characterize scenarios where operations on different isolated systems do not influence each other.

Next, we establish the connection between experimentally observable correlations and the maximum possible enhancement in predictive power achievable by any alternative theory. 
Our argument builds on the work of Barrett et al. \cite{barrett2006maximally}, which introduces a quantity of bipartite correlations formulated as chained Bell inequalities. 
Consider the previously described scenario and measurement setup, which yield the joint probability distribution \(P_{X Y \mid A B}\). To quantify the correlations, we define the following quantity \cite{barrett2006maximally}: 
\begin{align}\label{chainbell}
I_N& =I_N\left(  P_{X Y \mid A B}\right)\nonumber\\
&:=\sum_{i=1}^N\left(\left\langle\left[X_i-Y_i\right]\right\rangle+\left\langle\left[Y_i-X_{i+1}\right]\right\rangle\right) ,
\end{align}
where \(\langle \cdot \rangle=\sum_{k=0}^{d-1} k P(\cdot =k)\) denotes the average value of the random variable `` \(\cdot\) '', and \(\left[\cdot\right]\) denotes `` \(\cdot\) '' modulo \(d\). Defining \(X_{N+1}:=X_{1}+1\) modulo \(d\). 
Suppose that Alice and Bob share the maximally entangled state, \(\left|\psi_d\right\rangle=\frac{1}{\sqrt{d}} \sum_{j=0}^{d-1}|j\rangle_A|j\rangle_B\), and their measurements correspond to projections onto the states 
\begin{align}
   & |X_A\rangle=\frac{1}{\sqrt{d}} \sum_{j=0}^{d-1} \exp \left[\frac{2 \pi i}{d} j\left(X-\alpha_A\right)\right]|j\rangle\,,\label{setA}\\
   & |Y_B\rangle=\frac{1}{\sqrt{d}} \sum_{j=0}^{d-1} \exp \left[-\frac{2 \pi i}{d} j\left(Y-\beta_B\right)\right]|j\rangle\,,\label{setB}
\end{align}
where \(\alpha_A=(A-1/2)/N\) and \(\beta_B=B/N\). With these measurement settings, quantum mechanics predicts that \(I_N^{QM}=2\gamma/N+O\left(1 / N^2\right)\), where \(\gamma=\pi^2 /\left(4 d^2\right) \times\sum_{j=1}^{d-1} j / \sin ^2(\pi j / d)\). For sufficiently large \(N\), the quantity \(I_N^{QM}\) can be made arbitrarily small. 

We now present our main theorem, which demonstrates any nonsignaling model reproducing the correlations defined by \eqref{chainbell}-\eqref{setB} is inherently limited in its predictive power. 
To quantify the statistical distance between two \(d\)-dimensional probability distributions \(P_X\) and \(Q_X\), we define a quantity as \(\Delta\left(P_{X }, Q_{X }\right):= \sum_x\left|P_X(x)-Q_X(x)\right|/d\). 
\begin{theorem}\label{Theorem1}
For any non-signaling probability distribution, \(P_{XY Z \mid A B C}\), we have 
\begin{align}\label{lemmaeq}
 \Delta\left(P_{XZ|AC}, P_{\bar{X} }\times P_{Z \mid C}\right) 
   \leq \frac{d}{4}I_N\left(P_{X Y \mid A B}\right) ,
\end{align}
for all \(A, B,C\), and \(X\), where \(P_{\bar{X}}\) represent the uniform distribution on \(X\). The proof can be found in the Appendix \ref{proof}. 
\end{theorem}

\begin{figure}[t]
\centering
\includegraphics[width=8.6cm]{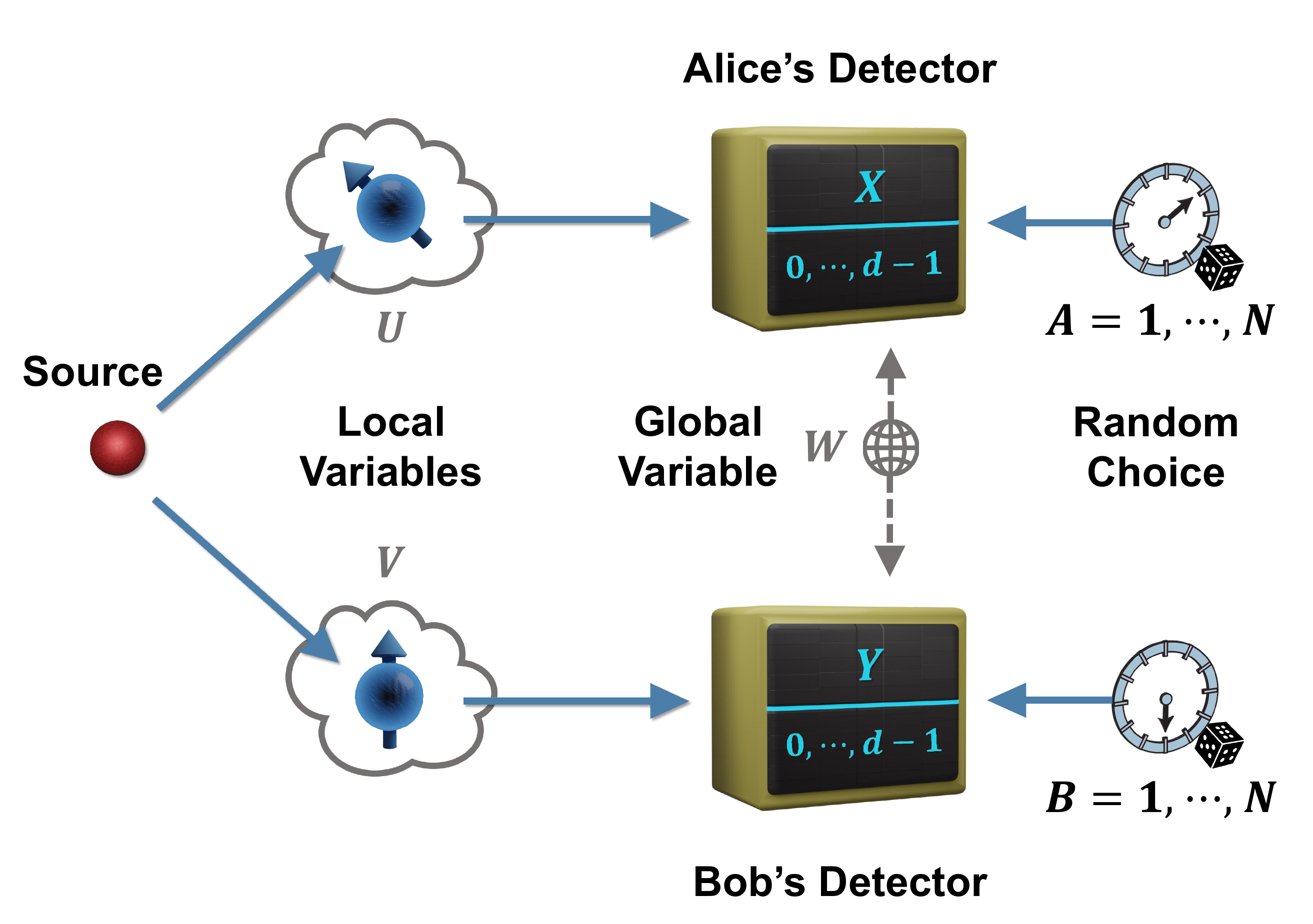}
\caption{{\sf A source emits two particles to spatially separated detectors controlled by Alice and Bob. Each party independently selects one of \(N\) measurements randomly, denoted \(A\) and \(B\), producing \(d\)-dimensional outcomes \(X\) and \(Y\). In hidden variable models, \(X\) and \(Y\) may depend on local variables \(U\) and \(V\), respectively, while \(W\) represents a nonlocal hidden variable influencing outcomes globally. }}
\label{fig1}
\end{figure}

The theorem is applicable to hidden variable theories, where the outcomes \(X\) and \(Y\) are determined by the measurement choices and hidden variables. Within the theoretical framework of Colbeck and Renner \cite{Colbeck}, hidden variables can be divided into local and nonlocal components. 
As illustrated in Fig. \ref{fig1}, we set \(U\) and \(V\) to denote Alice's and Bob's local variables, respectively, and \(W\) represents the nonlocal global variable. 
If ignoring \(W\), the experimental outcomes depend only on local parameters, with \(Z=(U, V)\) and \(C\) treated as a constant. 
The local variables are required to be physical, meaning they must not enable signaling between Alice and Bob. Formally, this condition is expressed as \(P_{XUV \mid ABC}=P_{XU \mid A}\) and \(P_{YUV \mid ABC}=P_{YV \mid B}\). 
Conversely, the nonlocal hidden variable \(W\) is not limited by the no-signaling condition, as signaling may become possible with knowledge of \(W\). The marginal distribution is defined as \(P_{X Y \mid ABUV}:=\sum_w P_W (w)P_{X Y \mid ABUVw}\), where \(w\) represents specific instances of the random variable \(W\). 
Assuming that measurement choices are independent, the condition \(P_{Z|AB}=P_Z\) holds for any fixed input pair \(\left ( A,B \right ) \). 
Using Theorem \ref{Theorem1}, the following bounds are derived 
\begin{align}
 \frac{4}{d}\Delta\left(P_{X U \mid A},  P_{\bar{X} }\times P_{U}\right) \leq  I_N \label{U}\\
\frac{4}{d}\Delta\left(P_{Y V \mid B}, P_{\bar{Y} }\times P_{V}\right) \leq I_N \label{V}
\end{align}
for all \(A\) and \(B\). 

Suppose enough rounds of experiment are conducted under the measurements scenario described in \eqref{chainbell}-\eqref{setB}, allowing us to determine an experimental optimal bound \(I^*_N\) on the quantity \(I_N\). 
According to \eqref{U}, the dependence of \(X\) on the local hidden variable \(U\), quantified as \(\frac{d}{4}\Delta\left(P_{X U \mid A},  P_{\bar{X} }\times P_{U}\right)\), is bounded by \(I^*_N\). A similar bound applies to \(Y\). 
For an ideal maximally entangled state and ideal experimental apparatus, quantum mechanics predicts that that \(I_N\propto 1/N\), which can be made arbitrarily small for sufficiently large \(N\). 
Thus, for any hidden variable model attempting to explain these quantum correlations, we conclude that \(P_{X U \mid A}=  P_{\bar{X} }\times P_{U}\) and \(P_{Y V \mid B}= P_{\bar{Y} }\times P_{V}\), implying that the measurement outcomes \(X\) and \(Y\) are virtually independent of the local hidden variables \(U\) and \(V\). This demonstrates that in arbitrary dimensions, certain quantum correlations cannot be explained by any hidden variable model with a nontrivial local component. 

\section{EXPERIMENTAL SETUP AND RESULTS}

\begin{figure}[tb]
\centering
\includegraphics[width=\columnwidth]{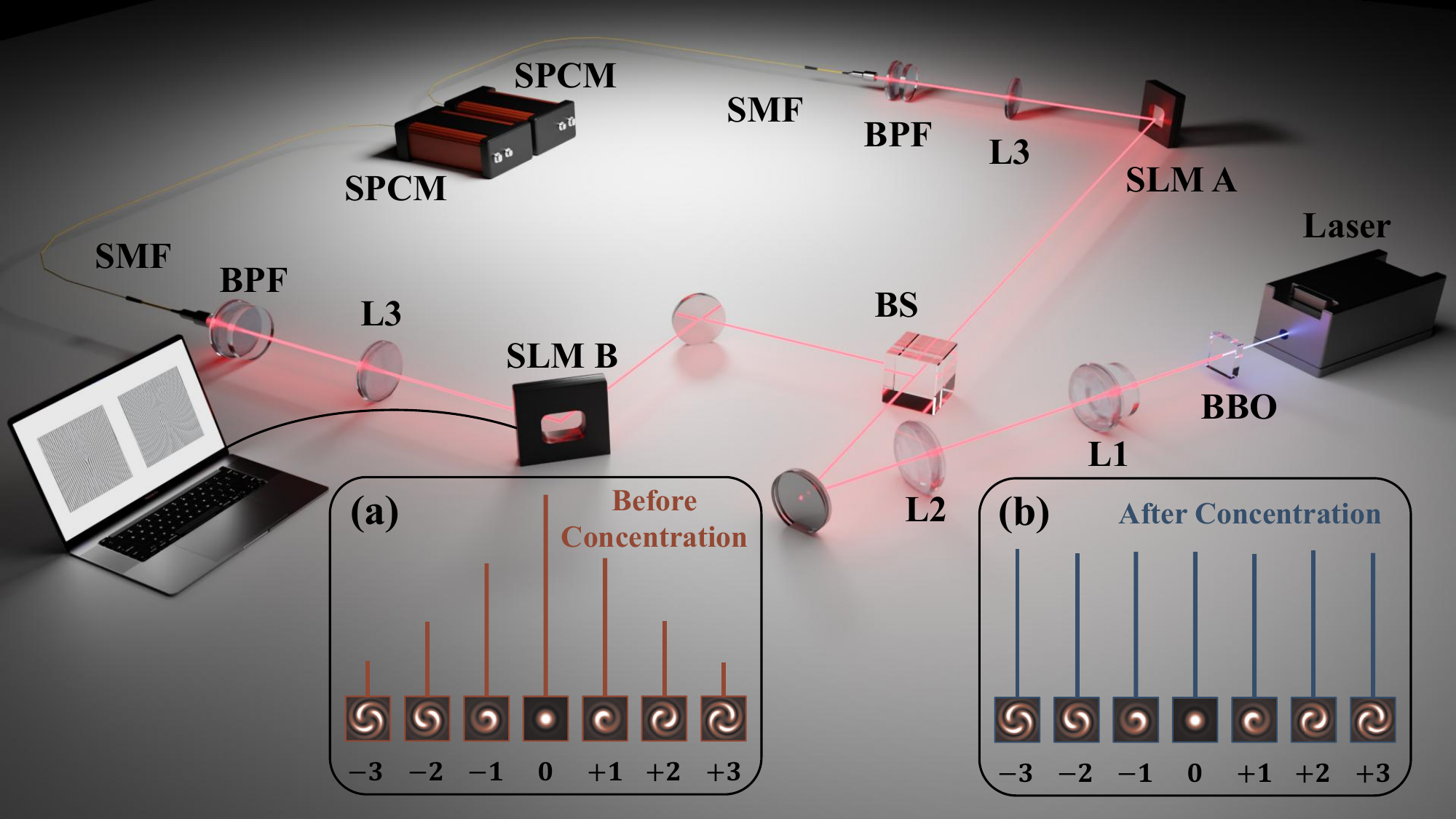}
\caption{{\sf Schematic diagram of the experimental setup. BBO \(\beta\)-barium borate crystal, BS 50:50 beam splitter, SLM spatial light modulator, L lens (\(f_1=100\) mm, \(f_2=400\) mm, \(f_3=500\) mm), BPF band-pass filter, SMF single-mode fiber, SPCM single photon counting module, the outputs are connected to a coincidence circuit. The inset (a) shows the original two-photon OAM spectrum of limited spiral bandwidth before entanglement concentration, while (b) shows the maximally entangled OAM spectrum after concentration, obtained by measuring the state \(\left | \ell  \right \rangle_{A}\otimes\left | -\ell  \right \rangle_{B}\).}}
\label{fig2}
\end{figure}

To establish the realistic experimental bounds of Theorem \ref{Theorem1}, we measure the correlation quantity \(I_N\) by utilizing the orbital angular momentum (OAM) entanglement. 
Distinguishing from polarization, which is naturally bidimensional, OAM is infinite-dimensional, such as a photon with \(\ell \) intertwined helical wavefronts, \(\left | \ell  \right \rangle\), carries \(\ell\hbar\) units of OAM \cite{mair2001entanglement}. Since these modes can define an infinitely dimensional discrete Hilbert space, and the number of effective dimensions can be readily tailored as required, this degree of freedom provides a practical route to entanglement with higher dimensions \cite{torres2003quantum}. 
Our experimental setup is shown in Fig. \ref{fig2}. We start by generating an entangled pair of photons through the SPDC process, by sending a mode-locked 355 nm Nd-YAG laser through a BBO crystal. We use a long-pass filter behind the crystal to block the pump beam, and then a 50:50 beam splitter (BS) separates the signal and idler photons. The conservation of orbital angular momentum ensures that if the signal photon is in the mode specified by $\left | \ell  \right \rangle$, the corresponding idler photon can only be in the mode $\left | -\ell  \right \rangle$~\cite{walborn2004entanglement}. Through this process, we generate photon pairs entangled according to the non-separable wavefunction \(\left | \Psi  \right \rangle =\sum_{\ell=-\infty }^{\ell=\infty}c_{\ell} \left | \ell  \right \rangle_{A}\otimes\left | -\ell  \right \rangle_{B}\), where \(c_{\ell}\) is the probability amplitude that the photon in arm \(A\) is in the state \(\left | \ell  \right \rangle_{A}\) and the photon in arm \(B\) is in \(\left | -\ell  \right \rangle_{B}\). 
In the detection, we use a combination of computer-controlled spatial light modulators (SLM) operating in reflection mode loaded with specially designed holographic gratings, single-mode fibers, and single-photon-counting module detectors to detect specific OAM quantum states. With appropriately designed phase holograms, an SLM can perform transforms of a photon in an arbitrary OAM superposition state to the Gaussian \(|\ell=0\rangle\) mode \cite{mair2001entanglement}. The hologram generation algorithm introduced in Ref. \cite{leach2005vortex} is applied to configure the SLMs. Then a 4-\textit{f} telescope consisting of two lenses is used to reimage the reflected photon and coupled into a single-mode fiber (SMF) which feeds an avalanche photon detector. An SMF projects the incoming photon into the fundamental mode of the fiber, a nearly Gaussian mode with \(m=p=0\). The count in the detector indicates the detection of the state in which the SLM was prepared. Band-pass filters (BF) with 10 nm width are placed in front of the SMF to reduce the detection of noise photons. The output of the detectors is connected to a coincidence circuit, the coincidence resolving time is 10 ns and an integration time of 30 s is used for the measurements.

\begin{figure}[t]
\centering
\includegraphics[width=8.6cm]{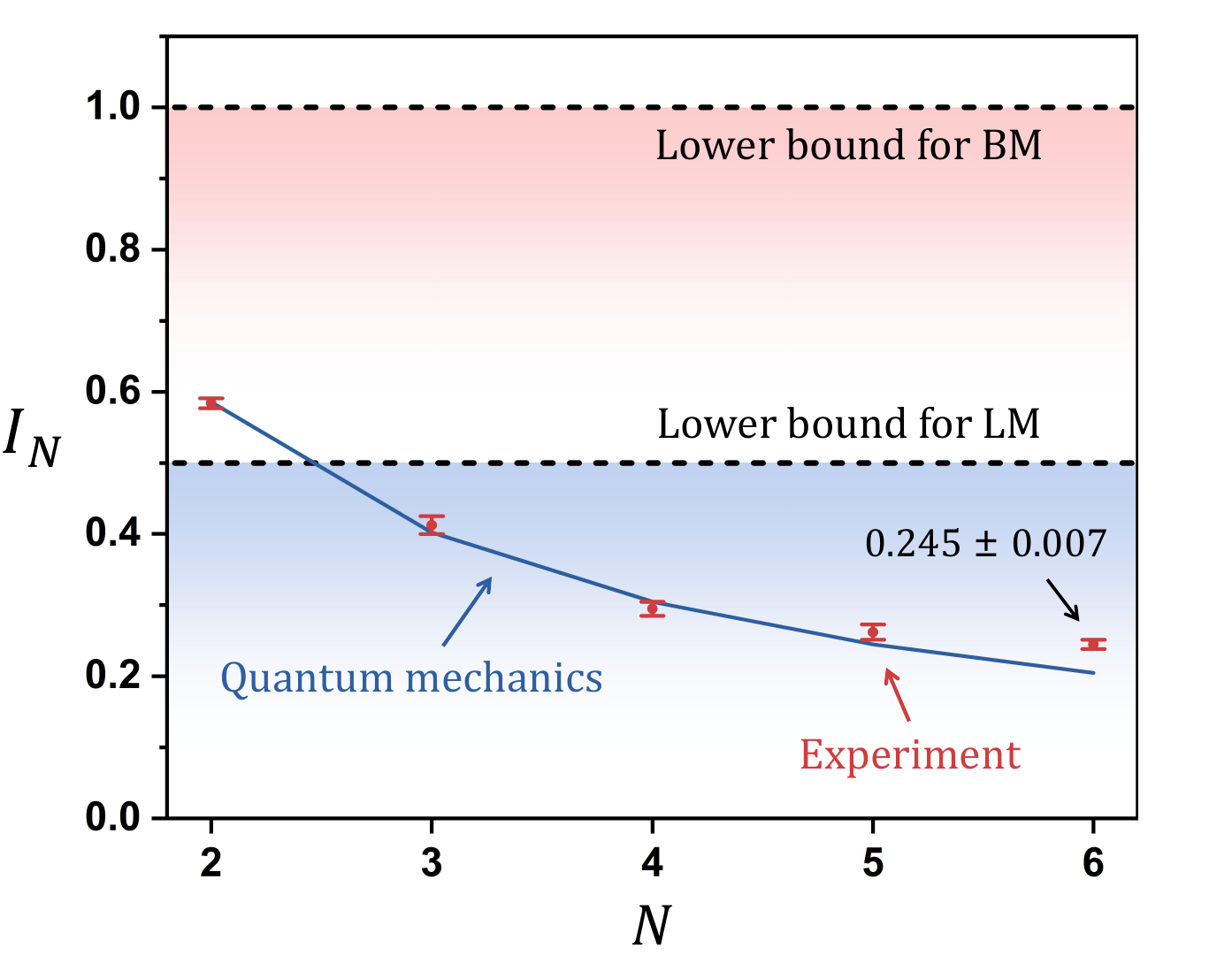}
\caption{{\sf The experimental value for the quantity \(I_N\) (red points with error bars) with a bi-dimensional maximally OAM entangled state, \(\left|\psi_2\right\rangle=\frac{1}{\sqrt{2}} \sum_{j=0}^{1}|j\rangle_A|j\rangle_B\), versus the number of measurement bases per side used, \(N\). 
The minimum value occurs at \(N=6\). 
The two dashed lines represent the lower bound restricted by Bell's model (BM) and Leggett's model (LM), respectively. 
The solid blue line indicates the theoretical curve predicted by quantum theory for the ideal case.}}
\label{fig3}
\end{figure}

The spiral bandwidth is an important factor affecting the degree of entanglement of the photon pairs for a given selected subspace of the OAM Hilbert space \cite{torres2003quantum}. As shown in Fig. \ref{fig2}(a), the finiteness of the spiral bandwidth leads to a non-maximally entangled state for the projection of the SPDC output state onto a \textit{d}-dimensional subspace. We perform the Procrustean filtering technique for entanglement concentration to enhance the entanglement \cite{bennett1996concentrating,law2004analysis}, which can be considered to be generalized measurements performed by local operations on the signal and idler beams. 
We performed local operations on the signal and idler photons, using alterations of the diffraction efficiencies of blazed phase gratings in the spatial light modulators to equalize amplitudes of OAM modes, to obtain a close approximation to a maximally entangled state, \(\left|\psi_d\right\rangle=\frac{1}{\sqrt{d}} \sum_{j=0}^{d-1}|j\rangle_A|j\rangle_B\), as shown in Fig. \ref{fig2}(b). 
The disadvantage of this method is the associated reduction in the number of detected photons~\cite{dada2011experimental}. Due to the experimental imperfections, such as optical misalignment, we adopt suitable OAM intervals $\Delta \ell$ to perform the measurements. Such a trade-off consideration enables us to obtain a relatively low crosstalk between orthogonal modes yet a relatively high coincidence count. For the tests within the subspaces of \(d=2\) to 4, we prioritized higher extinction between the orthogonal states while compensating for lower rates with longer integration times. We choose the modes \(\ell=-2, 2\) as the computational basis states for \(d=2\), and \(\ell=-3, 0, 3\) and \(\ell=-4, -1,1, 4\) for 3 and 4-dimension, respectively. 
For the test within the subspace of 5-dimension and 6-dimension, we choose the adjacent modes \(\ell=-2,-1,0,1,2\) and \(\ell=-3,-2,-1,1,2,3\) as the computational basis, respectively, for a higher count rate in fewer integration times.  

\begin{figure}[t]
\centering
\includegraphics[width=8.6cm]{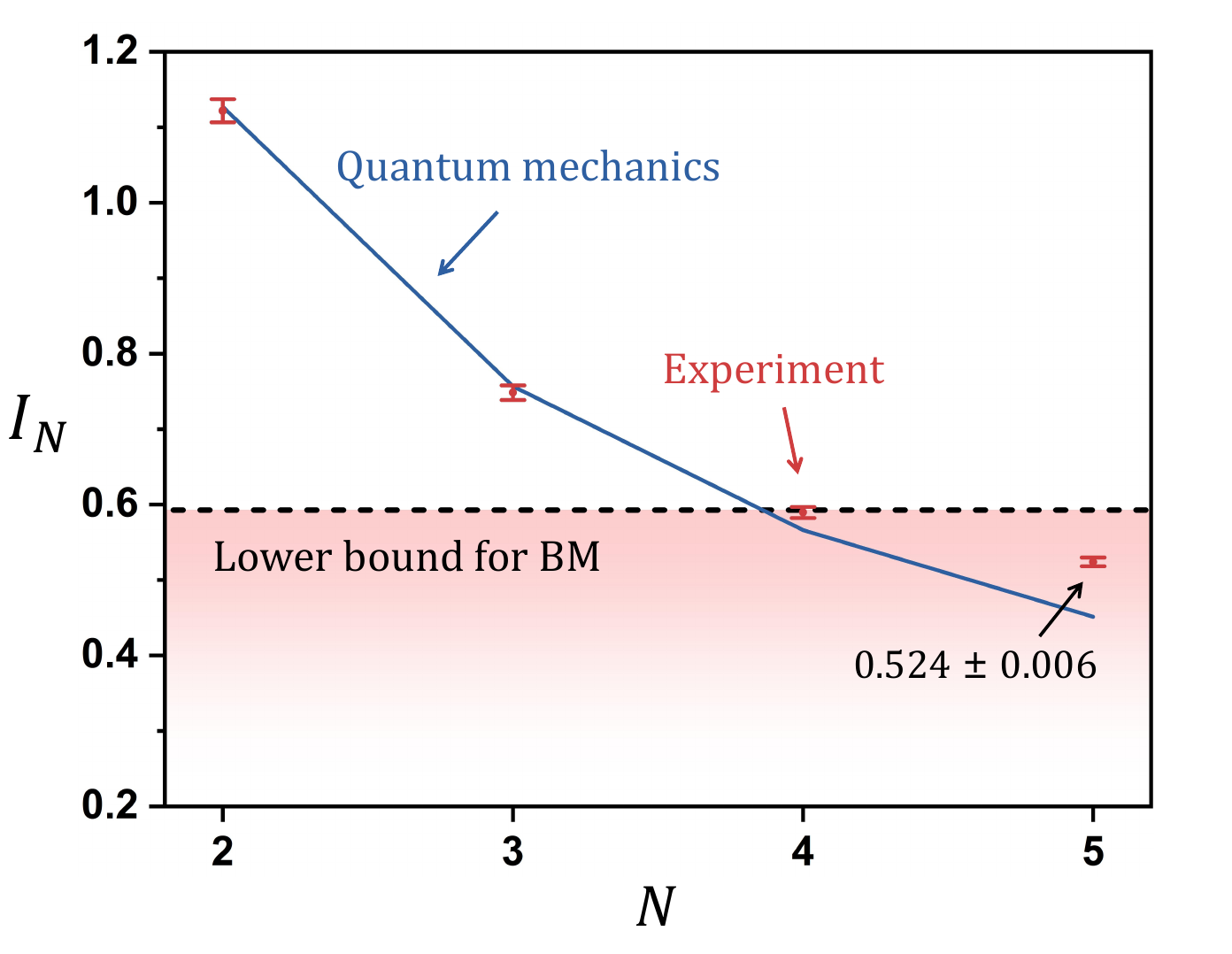}
\caption{{\sf The experimental value for the quantity \(I_N\) (red points with error bars) with a three-dimensional maximally OAM entangled state, \(\left|\psi_3\right\rangle=\frac{1}{\sqrt{3}} \sum_{j=0}^{2}|j\rangle_A|j\rangle_B\), versus the number of measurement bases per side used, \(N\). 
The minimum value occurs at \(N=5\). 
The dashed line represents the lower bound restricted by Bell's model (BM), which is clearly violated for all observations. 
The solid blue line indicates the theoretical curve predicted by quantum theory for the ideal case.}}
\label{fig4}
\end{figure}

Experimentally, we measured the correlation quantity \(I_N\) defined by \eqref{chainbell}. Distributing pairs of entangled particles between Alice and Bob and allowing each of them to perform a randomly chosen local measurement from \eqref{setA} and \eqref{setB}. 
By conducting rounds of measurements for various \(N\), we observe the quantity \(I_N\) reaches a minimum at a specific finite \(N\). This enables us to determine an optimal bound on the experimental minimum, denoted as \(I^*_N:=\min _N\left\{I_N\right\}\). 
In Fig.~\ref{fig3}, we measured \(I_N\) for the bidimensional OAM maximally entangled state, represented as \(\left|\psi_2\right\rangle=\frac{1}{\sqrt{2}} \sum_{j=0}^{1}|j\rangle_A|j\rangle_B\), and identified the minimum value, \(I^*_N=0.245\pm0.007\), occurring at \(N=6\). 
The experimental minimum cannot be optimized further for \(N>6\) as the correlation signal weakens and noise dominates. Achieving further lower values would require significant improvements in photon source quality and measurement precision. 
Above, we connected correlations \(I_N\) to the maximum independence of measurement outcomes \(X\) from local hidden variables \(U\), quantified by \( \frac{4}{d}\Delta\left(P_{X U \mid A},  P_{\bar{X} }\times P_{U}\right)\) in Eq. \eqref{U} (similar for \(Y\)). 
In Appendix \ref{limit}, we derive the lower bounds on \( I_N\) restricted by Bell's and Leggett's models from inequality \eqref{U}. For \(d=2\), the lower bound for Bell's model (BM) is \(1\), and for Leggett's model (LM) is \(0.5\). 
Our experimentally observed minimum value \(I^*_N=0.245\pm0.007\) violates the Bell model bound by 107 standard deviations and the Leggett model bound by 36 standard deviations, thereby falsifying both theories in the bidimensional system. 
In Fig. \ref{fig4}, we measured \(I_N\) for the three-dimensional OAM maximally entangled state, represented as \(\left|\psi_3\right\rangle=\frac{1}{\sqrt{3}} \sum_{j=0}^{2}|j\rangle_A|j\rangle_B\), and identified the minimum value, \(I^*_N=0.524\pm0.006\), occurring at \(N=5\). 
The lower bound on \(I_N\) restricted by Bell's model for three-dimensional systems is \(\frac{16}{27}\), and our experimental observation violates this bound by 11 standard deviations. 
Note that this bound differs from the bound \(d-1\) presented in \cite{barrett2006maximally}, despite being based on the same assumptions, as the inequalities \eqref{lemmaeq} were derived with a broader objective, sacrificing some tightness for dimensions \(d\) higher than \(2\). 

\begin{table}[t]
  \caption{ \label{tab1}
The table summarizes the experimental minimum value \(I^*_N\) for maximally entangled OAM states with different dimensions, ranging from \(d=2\) to \(d=6\). }

  \centering
  \renewcommand{\arraystretch}{1.5}
  \begin{ruledtabular}
        \begin{tabular}{ccc} 
        %\hline
        %\hline
        Entangled states&\(I^*_N\)&Experimental value\\ \hline 
        \(\left|\psi_2\right\rangle\)& \(I_{6}\)&$0.245\pm 0.007$\\ %\hline
        \(\left|\psi_3\right\rangle\)& \(I_{5}\)&$0.524\pm0.006$\\ %\hline
        \(\left|\psi_4\right\rangle\)& \(I_{5}\)&$0.835\pm0.013$\\ %\hline
        \(\left|\psi_5\right\rangle\)& \(I_{4}\)&$1.499\pm0.020$\\ %\hline
        \(\left|\psi_6\right\rangle\)& \(I_{3}\)&$2.429\pm0.042$\\ %\hline
        %\hline %\hline
        \end{tabular}
 \end{ruledtabular}
\end{table}

We further measured the quantity \(I_N\) for maximally entangled OAM states, with dimensions ranging from \(d=2\) to \(d=6\). 
By conducting multiple rounds of measurements for various \(N\), we identified the minimum values \(I^*_N\) for each dimension occurring at specific finite \(N\). 
In Table \ref{tab1}, we present the experimental minimum across various dimensional systems, establishing bounds on the potential increase in predictive power that any alternative theory could achieve. 
These bounds enable the falsification of a broader class of physical theories that provide significantly higher predictive accuracy than quantum theory, including those that remain theoretically ambiguous or experimentally unverified. 
Furthermore, if the number of independent measurement choices \(N\) is increased sufficiently large and the experimental apparatus is much improved, the value of \(I^*_N\) would approach zero, as predicted by quantum mechanics, thereby falsifying any alternative theories with increased predictive power than quantum theory.

\section{CONCLUSIONS}

In conclusion, we demonstrated that no extension can significantly surpass quantum mechanics in predictive accuracy for the local components of composite systems in arbitrary dimensions. 
This finding constitutes a further development of the non-extensibility theorem proposed by Colbeck and Renner \cite{Colbeck,colbeck2011no,colbeck2015completeness}, expanding its scope of application. 
By connecting experimentally observable correlations to the maximum potential increase in predictive power achievable by any alternative theory, we determined optimal experimental bounds across systems of varying dimensions. 
These bounds enable the falsification of a broader class of physical theories that provide significantly higher predictive accuracy than quantum theory, including Bell's and Leggett's models, and those that remain theoretically ambiguous or experimentally unverified. 
Our findings are expected to not only advance the foundational understanding of quantum theory but also pave the way for impactful applications in areas such as quantum cryptography, where the utilization of high-dimensional quantum states and the exploration of predictive power limits are of paramount importance.

\begin{acknowledgments} 
This work was supported by National Natural Science Foundation of China (12034016, 12205107), the National Key R\&D Program of China (2023YFA1407200), Natural Science Foundation of Fujian Province of China (2021J02002) for Distinguished Young Scientists (2015J06002), Program for New Century Excellent Talents in University (NCET-13-0495), Natural Science Foundation of Xiamen City (3502Z20227033), and Fundamental Research Funds for the Central Universities (ZQN-1206).
\end{acknowledgments}

\appendix
\section{PROOF OF THEOREM 1}\label{proof}

In this section, we prove Theorem \ref{Theorem1}, stated as Equation \eqref{lemmaeq} in the main text. The proof extends the argument in \cite{Colbeck,colbeck2011no}, and utilizes chained Bell inequalities from \cite{barrett2006maximally} as a quantification of the correlation.

\begin{proof}
From the fact that all probability distributions should be non-negative, we have 
\begin{align}\label{A1}
 \langle[\cdot]\rangle=\sum_{x=0}^{d-1} x P([\cdot]=x) \geq 1-P([\cdot]=0),
\end{align}
for any variable `` \(\cdot\) ''. We also have the following inequality under the no-signaling condition \cite{barrett2006maximally}: 
\begin{align}
 \begin{aligned}
&P\left(X_A=Y_B\right)=  \sum_{x=0}^{d-1} P\left(X_A=x, Y_B=x\right) \\
&\leq  \min \left(P\left(X_A=x\right), P\left(Y_B=x\right)\right)\\
&+\min \left(1-P\left(X_A=x\right), 1-P\left(Y_B=x\right)\right)\\
&=  1-\left|P\left(X_A=x\right)-P\left(Y_B=x\right)\right|.\label{A2}
\end{aligned}
\end{align}
where the function \(\min \left(P, P'\right)\) returns the smaller of the two values \(P\) and \(P'\). The inequality is valid for any certain \(x\in \{0, \ldots, d-1\}\). By applying the conditions derived in Eq. \eqref{A1} and Eq. \eqref{A2} to the correlations \(I_N\) in Eq. \eqref{chainbell}, and defining \(X_{N+1}:=X_{1}+1\pmod{d}\), we derive:   
\begin{align}
 &I_N\left(P_{X Y \mid A B  }\right):=\sum_{i=1}^N\left(\left\langle\left[X_i-Y_i\right]\right\rangle+\left\langle\left[Y_i-X_{i+1}\right]\right\rangle\right)\nonumber\\
 &\geq 2 N-\sum_{i=1}^N\left[P\left(X_i=Y_i\right)+P\left(X_{i+1}=Y_i\right)\right]\nonumber\\
 &\geq \sum_{i=1}^N[\left|P\left(X_i=x\right)-P\left(Y_i=x\right)\right|\nonumber\\
 &+\left|P\left(X_{i+1}=x\right)-P\left(Y_i=x\right)\right|]\nonumber\\
 &\geq \sum_{i=1}^N\left|P\left(X_i=x\right)-P\left(X_{i+1}=x\right)\right|\nonumber\\
 &\geq\left|P\left(X_i=x\right)-P\left(X_{i}=x+1\right)\right|,\label{A3}
\end{align}
for any \(x\in \{0, \ldots, d-1\}\). The first and second inequality arises from the conditions in Eq. \eqref{A1} and Eq. \eqref{A2}, the third and fourth follow from the triangle inequalities. 
In the subsequent analysis, we drop the subscript \(i\), which indicates the measurement choice, as the claim holds for any \(i\in \{1, \ldots, N\}\). 

If we hypothesize that \(P\left(X=x\right)>\frac{1}{d}+\frac{d}{4} I_N\), we can derive \(P\left(X=\left [x\pm p\right]\right)>\frac{1}{d}+(\frac{d}{4}-p ) I_N\) for any \(p\in \{0, \ldots,  \left \lfloor \frac{d}{2} \right \rfloor   \}\). Hence we can derive the following, 
\begin{align}\label{contra}
\sum_{x=0}^{d-1}P\left(X=x\right)&>1+\left(\frac{d^2}{4} -\sum_{q=0}^{d-1}\left [ q-\left \lfloor \frac{d}{2} \right \rfloor \right ] \right) I_N \nonumber\\
& = \begin{cases} 
1+\frac{1}{4}I_N, & \text{if } d \equiv 1 \mod{2} \\
1, & \text{if } d \equiv 0 \mod{2}
\end{cases}
\end{align}
which contradicts the normalization condition for probabilities, \(\sum_{x=0}^{d-1}P\left(X=x\right)=1\). Similarly, if we hypothesize that \(P\left(X=x\right)<\frac{1}{d}-\frac{d}{4} I_N\), it would imply \(\sum_{x=0}^{d-1}P\left(X=x\right)<1\), which is also a contradiction. By rejecting both hypotheses, we obtain 
\begin{align}\label{P1}
\left|P\left(X=x\right)-\frac{1}{d}\right|\leq\frac{d}{4} I_N,
\end{align}
for any \(x\in \{0, \ldots, d-1\}\). 

Considering potential alternatives to quantum theory that may introduce additional information, we evaluate \(I_N\) for the conditional distribution \(  P_{X Y Z\mid A BC}\) described in the main text, with fixed \(C\) and \(Z\). The following steps of this proof closely follow those in \cite{colbeck2011no}, which we repeat here for completeness. 
In this scenario, inequality \eqref{P1} implies that 
\begin{align}\label{PX}
\left|P_{xZ|ABC}-\frac{1}{d}\right|\leq\frac{d}{4} I_N\left(P_{X Y Z\mid A BC}\right) ,
\end{align}
for any \(A, B,C\), and \(x\). 
We now average both sides of \eqref{PX} over \(z\), where \(z\) represents specific instances of the random variable \(Z\). 
The right-hand-side gives 
\begin{align}
&\sum_z P_{Z \mid ABC}(z) I_N\left(P_{X Yz\mid A BC}\right)\nonumber\\
&=\sum_z P_{Z \mid C}(z) I_N\left(P_{X Yz \mid A BC}\right) \nonumber\\
& =I_N\left(P_{X Y \mid A BC}\right),
\end{align}
where we used the non-signaling condition \(P_{Z \mid A B C}  =P_{Z \mid C}\), which is implied by \eqref{XZ} and \eqref{YZ}. 
Then, taking the average on the left-hand side of \eqref{PX} gives \(\sum_z P_{Z \mid ABC}(z) \left|P_{xz|ABC}-\frac{1}{d}\right|=\left|P_{xZ|ABC}-\frac{1}{d} P_{Z \mid ABC}\right|\). Therefore, we have 
\begin{align}\label{A8}
   & \left|P_{xZ|ABC}-\frac{1}{d} P_{Z \mid ABC}\right| \nonumber\\
   &\leq \frac{d}{4}I_N\left(P_{X Y \mid A B C}\right)
    =\frac{d}{4}I_N\left(P_{X Y \mid A B}\right),
\end{align}
where the last equality follows from the non-signaling condition \eqref{XY}. 
Note that inequality \eqref{A8} holds for all \(x\). Let \(P_{\bar{X}}\) represent the uniform distribution on \(X\), such that \(P_{\bar{X}}(x)=\frac{1}{d}\) for all \(x\). Using our definition of the statistical distance \(\Delta\left(P_{X }, Q_{X }\right):= \sum_x\left|P_X(x)-Q_X(x)\right|/d\), we have that 
\begin{align}
& {\sum_x\left|P_{xZ|ABC}-\frac{1}{d} P_{Z \mid ABC}\right|}/d \nonumber\\
&=\Delta\left(P_{XZ|ABC}, P_{\bar{X} }\times P_{Z \mid ABC}\right) \nonumber\\
   &\leq \frac{d}{4}I_N\left(P_{X Y \mid A B}\right),
\end{align}
Using the property that the variational distance of marginal distributions cannot exceed that of the joint distributions, i.e., \(\Delta\left(P_{X }, Q_{X }\right)\leq \Delta\left(P_{X Y}, Q_{X Y}\right)\) for any \(P_{X Y}\) and \(Q_{X Y}\), we conclude 
\begin{align}\label{A9}
   \Delta\left(P_{XZ|AC}, P_{\bar{X} }\times P_{Z \mid C}\right) 
   \leq \frac{d}{4}I_N\left(P_{X Y \mid A B}\right).
\end{align}
This completes the proof of Theorem \ref{Theorem1}. 

\end{proof}

\section{LIMITATION ON HIDDEN VARIABLE THEORIES}\label{limit}

In the main text, we connected experimentally observable correlations \(I_N\) to the maximum independence of measurement outcomes \(X\) from local hidden variables \(U\), quantified by \( \frac{4}{d}\Delta\left(P_{X U \mid A},  P_{\bar{X} }\times P_{U}\right)\) in Eq. \eqref{U} (similar for \(Y\)). 
In this section, we review well-known hidden variable theories, including Bell's model and Leggett's model. For each framework, we derive lower bounds on \( I_N\), representing the constraints these models impose on the observed correlations.

In the Bell model \cite{bell1987speakable}, it is postulated that the measurement outcomes of Alice and Bob are fully determined by local parameters. Within the bipartite setup described above, this means that Alice's outcome \(X\) is described by a deterministic function \(X=f(A, U)\), where \(A\) represents the measurement setting, and \(U\) represents the local hidden variable on this side. Similarly, Bob's outcome \(Y\) is given by \(Y=g(B, V)\). This assumption reflects the principle of locality, which implies that the outcomes are independent of any non-local variables or the distant measurement setting. 
Determinism and local causality together imply that all probabilities \(P_{X }(x)\) and \(P_{Y }(y)\) must be equal to either \(0\) or \(1\). 
By applying our previous analysis, where \(C\) is treated as a constant, with the normalization condition for probabilities \(\sum_{x=0}^{d-1}P_{X }(x)=1\), we obtain 
\begin{align}
& \Delta\left(P_{X U \mid A},  P_{\bar{X} }\times P_{U}\right)\geq\Delta\left(P_{X  \mid A},  P_{\bar{X} }\right) \nonumber\\
& = {\sum_{x=0}^{d-1}\left|P_{X}(x)-\frac{1}{d} \right|}/d\nonumber\\
& = \frac{2(d-1)}{d^2},
\end{align}
where the first inequality arises from the property that \(\Delta\left(P_{X }, Q_{X }\right)\leq \Delta\left(P_{X Y}, Q_{X Y}\right)\). Thus, the inequality \eqref{U} can be expressed as 
\begin{align}
    I_N \geq \frac{8(d-1)}{d^3}   ,
\end{align}
which implies that if we measured the quantity \(I_N\) is less than the value \(\frac{8(d-1)}{d^3}\), the observed quantum correlations cannot be explained by the Bell model. 
Note that this bound differs from the bound \(d-1\) presented in \cite{barrett2006maximally}, despite being based on the same assumptions, as the inequalities \eqref{lemmaeq} were derived with a broader objective, sacrificing some tightness for dimensions \(d\) higher than \(2\). 

The fundamental assumption of Leggett’s model is that, at the local level, each individual quantum system behaves as if it is always in a pure state. On each side, the subsystem is described by bi-dimensional pure states \(|\vec{u}\rangle\) and \(|\vec{v}\rangle\), which are represented by three-dimensional unit vectors \(\vec{u}\) and \(\vec{v}\) on the Poincar\'e sphere. 
Note that Leggett's framework is restricted to bi-dimensional systems, as it specifies only the outcomes for measurements on spin-half particles. 
To integrate this framework into our general discussion, these vectors can be interpreted as the local components of hidden variables, \(U\) and \(V\). 
As above, the measurement choices on each side are denoted by \(A\) and \(B\), respectively. When restricted to projective spin measurements, these choices can be represented by three-dimensional vectors \(\vec{a}\) and \(\vec{b}\), which indicate their spatial orientation. 
The local marginal probability distributions follow to the spin-projection rule, given as 
\begin{align}\label{model}
P_{X \mid AU}(x) & =\frac{1}{2}\left[1+(-1)^x \vec{a} \cdot \vec{u}\right], \\
P_{Y \mid BV}(y) & =\frac{1}{2}\left[1+(-1)^y \vec{b} \cdot \vec{v}\right] ,
\end{align}
which, for a photonic spin-half system, aligns with the well-known Malus' law for polarization. 
By applying our previous analysis, where \(C\) is treated as a constant, we obtain 
\begin{align}
& \Delta\left(P_{X U \mid A},  P_{\bar{X} }\times P_{U}\right) =\left \langle \Delta\left(P_{X \mid \vec{a},\vec{u}},  P_{\bar{X} }\right) \right \rangle_{\vec{u}}  \nonumber\\
& = \left \langle \frac{1}{2}\sum_{x=0}^{1}\left|P_{X|\vec{a},\vec{u}}(x)-\frac{1}{2} \right| \right \rangle_{\vec{u}}\nonumber\\
& = \frac{\left \langle \left|\vec{a} \cdot \vec{u} \right| \right \rangle_{\vec{u}} }{2}\,.
\end{align}
Thus the inequality \eqref{U} implies that 
\begin{align}
    I_N \geq \left \langle \left|\vec{a} \cdot \vec{u} \right| \right \rangle_{\vec{u}} \, .
\end{align}
Since the model is not fully specified due to the lack of an explicit distribution of the hidden variable \(\vec{u}\), we analyze the model under two possible cases, considering the implications of different distributions for \(\vec{u}\). 
First, consider the case where the vector \(\vec{u}\) is fixed in a specific direction within the same plane on the Bloch sphere as the measurement settings. 
In this case, \(\left|\vec{a} \cdot \vec{u} \right|\leq\cos \frac{\pi}{2 N}\) , and the maximum value is achieved when the vector \(\vec{u}\) is positioned exactly midway between the two settings of \(\vec{a}\). The inequality becomes 
\begin{align}
    I_N \geq \cos \frac{\pi}{2 N} \, .
\end{align}
If the vector \(\vec{u}\) is no longer confined to the plane of measurements, we can introduce a second set of measurements, identical to the first except that it is contained in an orthogonal plane. Such that \(\left|\vec{a} \cdot \vec{u} \right|\leq\frac{1}{\sqrt{2} } \cos \frac{\pi}{2 N}\). 
Next, consider a modified case where \(\vec{u}\) is uniformly distributed over the entire Bloch sphere, which is a more natural assumption compared to restricting it to a specific direction corresponding to the chosen measurements. 
In this case, the inequality becomes  
\begin{align}\label{LMbound}
    I_N \geq \int_{\theta=0}^\pi \mathrm{d} \theta \frac{|\cos \theta| \sin \theta}{2}=\frac{1}{2}   ,
\end{align}
where \(\theta\) is defined as the angle between \(\vec{a}\) and \(\vec{z}\). Equation \eqref{LMbound} implies that if the measured value of quantity \(I_N\) is less than \(\frac{1}{2}\), the observed quantum correlations cannot be explained by the Leggett models.

\bibliographystyle{apsrev4-2} 

\end{document}